\documentclass{PoS}

\title{The $\mathbf{B_s \to \mu^+\mu^-}$ and $\mathbf{B_d \to \mu^+\mu^-}$ Decays: \\ Standard Model and Beyond}
\ShortTitle{The $B_{s,d} \to \mu^+\mu^-$ Decays: Standard Model and Beyond}

\author{\speaker{Wolfgang Altmannshofer}
        \\
        Fermi National Accelerator Laboratory, P.O. Box 500, Batavia, IL 60510, USA\\
        E-mail: \email{waltmann@fnal.gov}}

\abstract{I review the status of the rare decays $B_s \to \mu^+\mu^-$ and $B_d \to \mu^+\mu^-$. The latest Standard Model predictions for the corresponding branching ratios are discussed.
An overview of additional observables that are accessible in the $B_s \to \mu^+\mu^-$ decay is given. 
The new physics implications of the recent measurement of the $B_s \to \mu^+\mu^-$ branching ratio are discussed both model independently and in the context of the Minimal Supersymmetric Standard Model.}

\FullConference{14th International Conference on B-Physics at Hadron Machines,\\
		April 8-12, 2013\\
		Bologna, Italy }

\begin{document}

\section{Introduction}

Recently, the LHCb collaboration announced first evidence for the rare decay $B_s \to \mu^+\mu^-$~\cite{Aaij:2012nna}. Using $2.1$fb$^{-1}$ of data, an excess of $B_s \to \mu^+\mu^-$ candidates, $3.5\sigma$ above background, was observed.
Performing a fit to the signal leads to a best fit value for the branching ratio of~\cite{Aaij:2012nna}
\begin{equation} \label{eq:BR_exp}
{\rm BR}(B_s \to \mu^+ \mu^-)_{\rm exp} = (3.2^{+1.5}_{-1.2}) \times 10^{-9} ~.
\end{equation}
This value is in excellent agreement with the Standard Model (SM) prediction, discussed below.
Limits on the $B_d \to \mu^+\mu^-$ branching ratio are still almost an order of magnitude above the corresponding SM prediction. The current most stringent bound is set by LHCb and reads~\cite{Aaij:2012nna}
\begin{equation} 
{\rm BR}(B_d \to \mu^+ \mu^-)_{\rm exp} < 9.4 \times 10^{-10} ~@~ 95\% ~{\rm C.L.}~.
\end{equation}
The tiny branching ratios of these decays in the SM are due to several factors: (i) loop suppression, (ii) GIM suppression, and (iii) helicity suppression. Extensions of the SM do not necessarily contain any of these suppression mechanisms. Therefore, $B_s \to \mu^+\mu^-$ and $B_d \to \mu^+\mu^-$ are highly sensitive probes of the flavor sector of models of new physics (NP).

In this talk I give a short theory review of the rare decays $B_s \to \mu^+\mu^-$ and $B_d \to \mu^+\mu^-$ in the SM and beyond. In section~\ref{sec:SM}, I discuss the latest SM predictions for the branching ratios and give an overview of additional observables that can be accessed in the $B_s \to \mu^+\mu^-$ decay. In section~\ref{sec:NP}, I discuss the implications of the recent experimental results for NP, both model independently and in the context of the Minimal Supersymmetric Standard Model.

\section{The $\mathbf{B_s \to \mu^+\mu^-}$ and $\mathbf{B_d \to \mu^+\mu^-}$ Decays in the Standard Model} \label{sec:SM}

The $B_{s,d} \to \mu^+\mu^-$ decays can be described by an effective Hamiltonian (see e.g.~\cite{Altmannshofer:2012az})
\begin{equation} \label{eq:Heff}
 \mathcal{H}_{\rm eff} = - \frac{4 G_F}{\sqrt{2}} V_{tb} V_{ts}^* \frac{e^2}{16\pi^2} \sum_i \Big( C_i \mathcal{O}_i + C_i^\prime \mathcal{O}_i^\prime \Big) ~,
\end{equation}
that consists of flavor changing dimension 6 operators $\mathcal{O}_i^{(\prime)}$ and their corresponding Wilson coefficients $C_i^{(\prime)}$.
In the SM, the only non-negligible Wilson coefficient entering the predictions for $B_{s,d} \to \mu^+\mu^-$ is $C_{10}$. This Wilson coefficient is known in the SM to very high precision. NLO QCD corrections have been computed in~\cite{Misiak:1999yg,Buchalla:1998ba}; NLO electro-weak corrections are know in the large top mass limit~\cite{Buchalla:1997kz}. The leftover electro-weak renormalization scheme dependence of $C_{10}$ leads to the largest {\it intrinsic theory uncertainty} in the SM prediction of the $B_{s,d} \to \mu^+\mu^-$ branching ratios of about 5\%~\cite{Buras:2012ru}. This uncertainty could be reduced by a full 2-loop calculation of the electro-weak corrections.
Thanks to impressive progress on the lattice~\cite{Dowdall:2013tga}, the main {\it parametric uncertainty} in the SM predictions for the branching ratios are not anymore the $B$ meson decay constants $f_{B_d}$ and $f_{B_s}$. The parametric uncertainties are now dominated by the relevant CKM matrix elements. 
The most recent theory predictions for the SM branching ratios read
\begin{eqnarray}
 {\rm BR}(B_s \to \mu^+\mu^-)_{\rm SM} &=& (3.25 \pm 0.17) \times 10^{-9} ~~~\cite{Buras:2013uqa}~, \label{eq:Bsmumu_SM}\\
 {\rm BR}(B_d \to \mu^+\mu^-)_{\rm SM} &=& (1.03 \pm 0.07) \times 10^{-10} ~, \label{eq:Bdmumu_SM}
\end{eqnarray}
where the error estimates are purely parametric. The above SM prediction for $B_d \to \mu^+\mu^-$ updates the value in~\cite{Buras:2012ru}, by using the latest lattice value for $f_{B_d} = 186(4)$~MeV~\cite{Dowdall:2013tga}.

The $B_{s,d} \to \mu^+\mu^-$ decays are inevitably accompanied by additional photon emission. The dominant correction arises from bremsstrahlung radiation, and the predictions in~(\ref{eq:Bsmumu_SM}) and (\ref{eq:Bdmumu_SM}) refer to the branching ratios fully inclusive of bremsstrahlung~\cite{Buras:2012ru}. Corrections due to direct emission of soft photons are well below the percent level and therefore negligible~\cite{Buras:2012ru,Aditya:2012im}.

Measured in experiments are time-integrated untagged rates, while in the above theory predictions, the effect of meson oscillations is not taken into account. Therefore, when comparing the SM prediction for $B_s \to \mu^+\mu^-$ with experimental results, one has to account for the sizable width difference $y_s = \tau_{B_s} \Delta \Gamma_s/2 = (8 \pm 1)\%$ in the $B_s$ system~\cite{Aaij:2013oba} (see \cite{DeBruyn:2012wj,DeBruyn:2012wk,Rob}):
\begin{equation}
 \overline{\rm BR}(B_s \to \mu^+\mu^-)_{\rm SM} = \frac{1 + y_s \mathcal{A}_{\Delta \Gamma}^{\mu\mu} }{1-y_s^2} {\rm BR}(B_s \to \mu^+\mu^-)_{\rm SM} ~.
\end{equation}
The mass eigenstate rate asymmetry $\mathcal{A}_{\Delta \Gamma}^{\mu\mu} \in [-1,1]$ depends in general on NP. In the SM, $\mathcal{A}_{\Delta \Gamma}^{\mu\mu}  = +1$ and the corresponding rescaled SM prediction, that can be directly compared to the experimental value, reads~\cite{Buras:2013uqa}
\begin{equation}
 \overline{\rm BR}(B_s \to \mu^+\mu^-)_{\rm SM} = \frac{1}{1-y_s} {\rm BR}(B_s \to \mu^+\mu^-)_{\rm SM} = (3.56 \pm 0.18) \times 10^{-9} ~.
\end{equation}
This value is in remarkably good agreement with the experimental measurement~(\ref{eq:BR_exp}).
The small finite width difference in the $B_d$ system has negligible impact on BR$(B_d \to \mu^+\mu^-)$.

The mass eigenstate rate asymmetry $\mathcal{A}_{\Delta \Gamma}^{\mu\mu}$ can be determined experimentally, by performing a measurement of the $B_s \to \mu^+\mu^-$ effective lifetime~\cite{DeBruyn:2012wk}
\begin{equation}
\tau_{\mu\mu} = \frac{\int_0^\infty dt~t~\langle \Gamma(B_s(t) \to \mu^+\mu^-)\rangle}{\int_0^\infty dt ~ \langle \Gamma(B_s(t) \to \mu^+\mu^-)\rangle} = \frac{\tau_{B_s}}{1-y_s^2} \left( \frac{1 + 2 \mathcal{A}_{\Delta \Gamma}^{\mu\mu} y_s + y_s^2}{1 + \mathcal{A}_{\Delta \Gamma}^{\mu\mu} y_s} \right) ~.
\end{equation}
A high precision measurement of $\tau_{\mu\mu}$, which is required to extract $\mathcal{A}_{\Delta \Gamma}^{\mu\mu}$, might be feasible with the large data samples expected from an LHCb upgrade~\cite{Bediaga:2012py}. 
A measurement of $\mathcal{A}_{\Delta \Gamma}^{\mu\mu}$ could reveal NP effects, even if the $B_s \to \mu^+\mu^-$ branching ratio is close to the SM prediction.

The full time-dependent flavor-tagged $B_s \to \mu^+\mu^-$ decay rate would allow to measure in addition also a CP asymmetry~\cite{DeBruyn:2012wk,Buras:2013uqa}
\begin{equation}
 \frac{\Gamma(B_s(t) \to \mu^+\mu^-) - \Gamma(\bar B_s(t) \to \mu^+\mu^-)}{\Gamma(B_s(t) \to \mu^+\mu^-) + \Gamma(\bar B_s(t) \to \mu^+\mu^-)} = \frac{S_{\mu\mu} \sin(\Delta M_s t)}{\cosh(\Gamma_s y_s t) + \mathcal{A}^{\mu\mu}_{\Delta \Gamma} \sinh(\Gamma_s y_s t)} ~.
\end{equation}
The quantity $S_{\mu\mu}$ is sensitive to CP violation both in the $B_s \to \mu^+\mu^-$ decay amplitude and in $B_s$ mixing.
It would give interesting complementary information on possible CP violating NP contributions to $B_s \to \mu^+\mu^-$, even if the branching ratio turns out to be SM-like to a high precision.
Given the expected statistics, a measurement of $S_{\mu\mu}$ will be challenging even with a LHCb upgrade.

\section{The $\mathbf B_{s,d} \to \mu^+\mu^-$ Decays and New Physics}  \label{sec:NP}

Due to the high sensitivity to NP, the recent measurement of the $B_s \to \mu^+\mu^-$ branching ratio leads to strong constraints on the flavor sector of extensions of the SM. I discuss such constraints both model independently and in the Minimal Supersymmetric extension of the SM.

\subsection{Model Independent Bounds on New Physics from $\mathbf{B_s \to \mu^+\mu^-}$}

While in the SM only the Wilson coefficient $C_{10}$ is relevant for the description of the $B_s \to \mu^+\mu^-$ decay,
in extensions of the SM, also scalar and pseudo-scalar Wilson coefficients ($C_S$ and $C_P$) as well as the corresponding ``right-handed'' coefficients ($C_{10}^\prime$, $C_S^\prime$, and $C_P^\prime$) can become relevant (see e.g.~\cite{Altmannshofer:2012az} for a definition of the corresponding operators in the effective Hamiltonian~(\ref{eq:Heff}))
\begin{equation} 
{\rm BR}(B_s \to \mu^+ \mu^-) \propto m_\mu^2 \left( \left| (C_{10}^{\rm SM} + C_{10}^{\rm NP} - C_{10}^\prime) + \frac{m_{B_s}}{2m_\mu} (C_P - C_P^\prime) \right|^2 + \left| \frac{m_{B_s}}{2m_\mu} ( C_S - C_S^\prime ) \right|^2 \right) ~.
\end{equation}
Note that in the presence of the scalar and pseudo-scalar coefficients, the helicity suppression of the branching ratio by the muon mass is lifted and strong constraints on $C_S^{(\prime)}$ and $C_P^{(\prime)}$ can be obtained.

\begin{figure}[tb] \centering
\includegraphics[width=.46\textwidth]{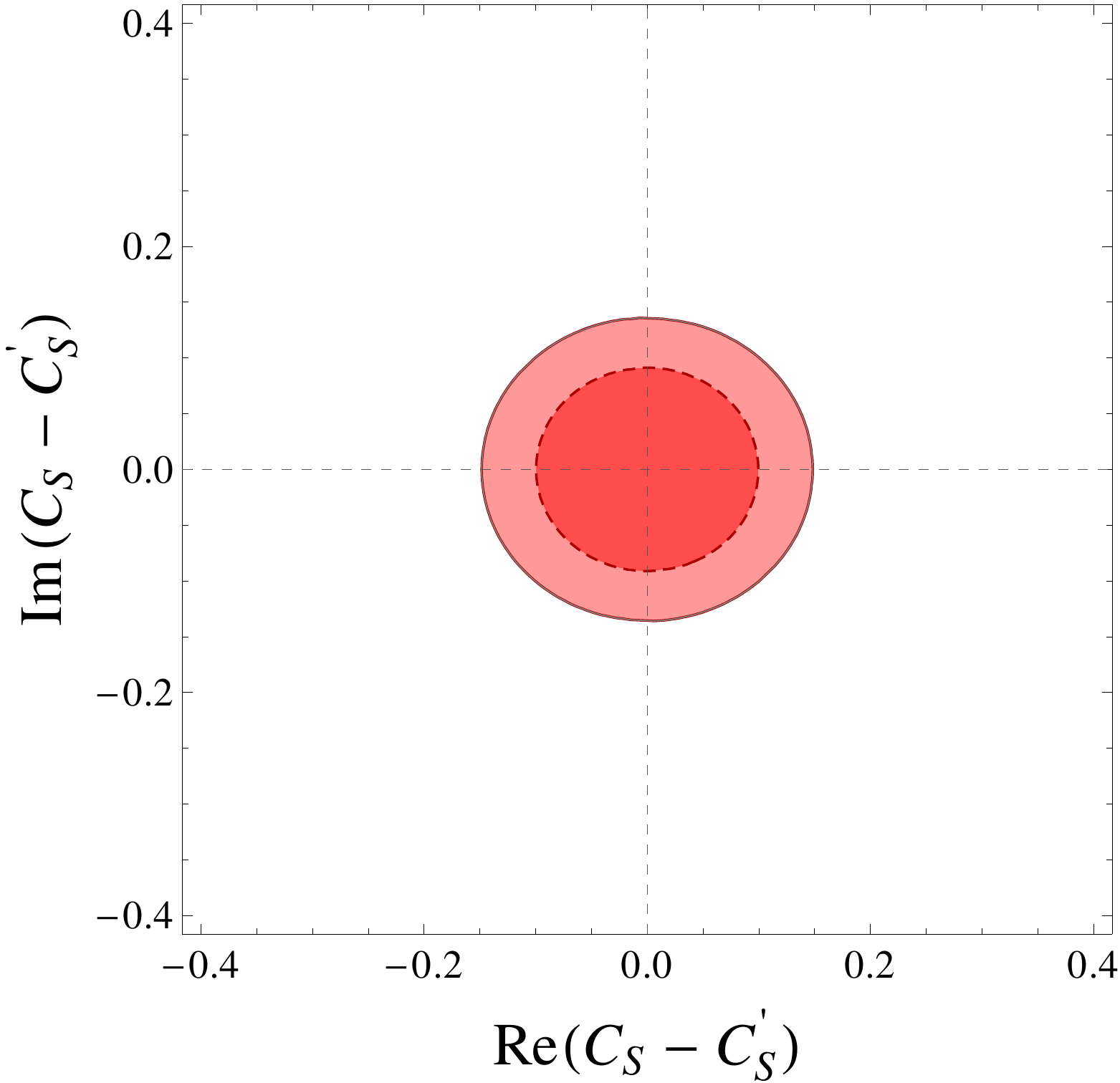} ~~~~~~
\includegraphics[width=.46\textwidth]{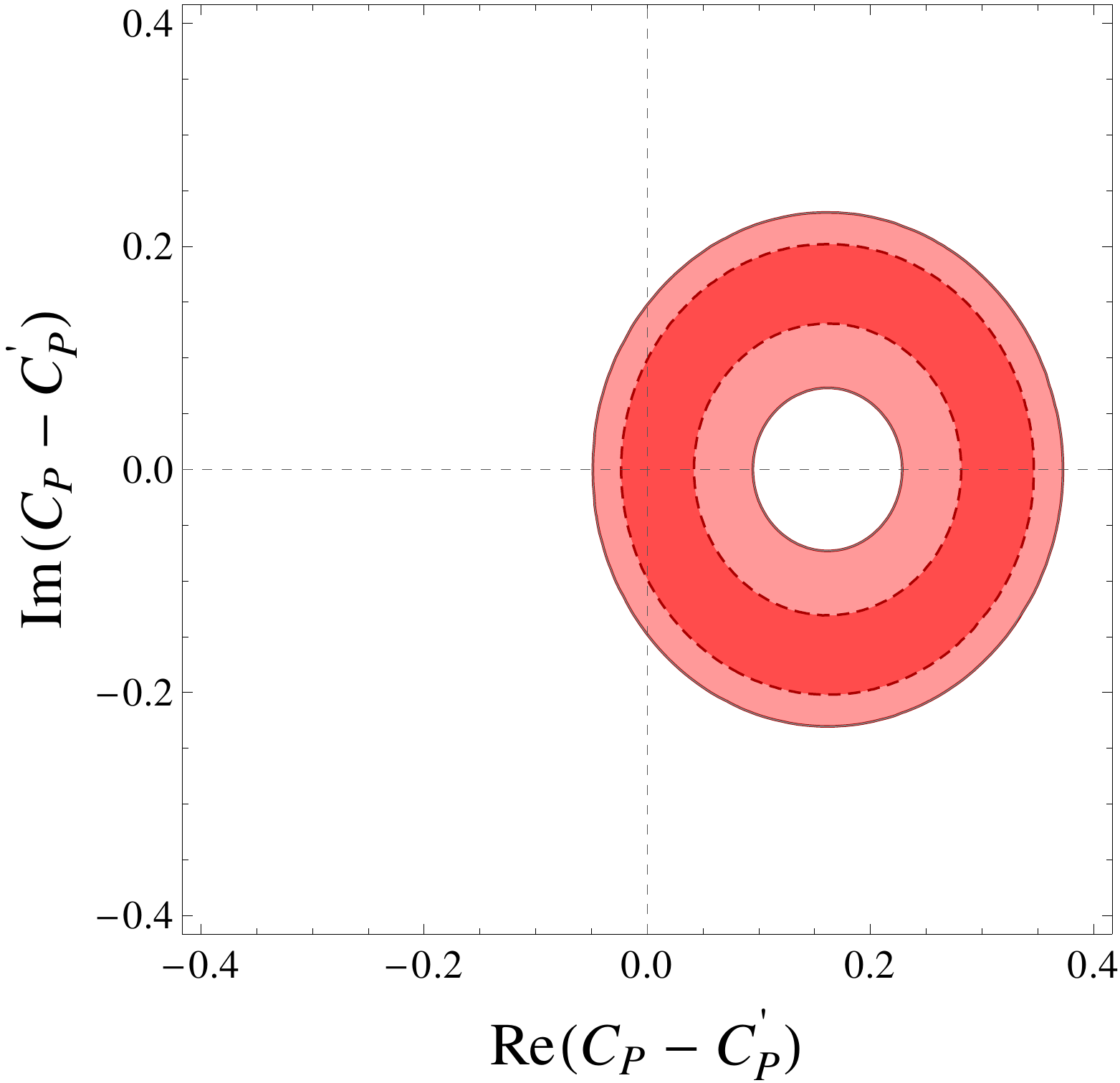}
\caption{Constraints on the scalar Wilson coefficients $C_S - C_S^\prime$ (left plot) and $C_P - C_P^\prime$ (right plot) at $1\sigma$ (dark red) and $2\sigma$ (light red) from BR$(B_s \to \mu^+ \mu^-)$, assuming no new physics in $C_{10} - C_{10}^\prime$. (Update of~\cite{Altmannshofer:2012az}.)}
\label{fig:bounds}
\end{figure}

Figure~\ref{fig:bounds}, shows the constraints on the relevant combinations $C_S - C_S^\prime$ (left plot) and $C_P - C_P^\prime$ (right plot) from the measurement of BR$(B_s \to \mu^+ \mu^-)$, assuming only NP in the scalar or pseudo-scalar coefficients, respectively. Constraints on the orthogonal combinations $C_S + C_S^\prime$ and $C_P + C_P^\prime$ can be obtained from measurements of $B \to K \mu^+\mu^-$~\cite{Becirevic:2012fy}. The corresponding constraints are generically weaker by one order of magnitude.

\begin{figure}[tb]
\centering
\includegraphics[width=.46\textwidth]{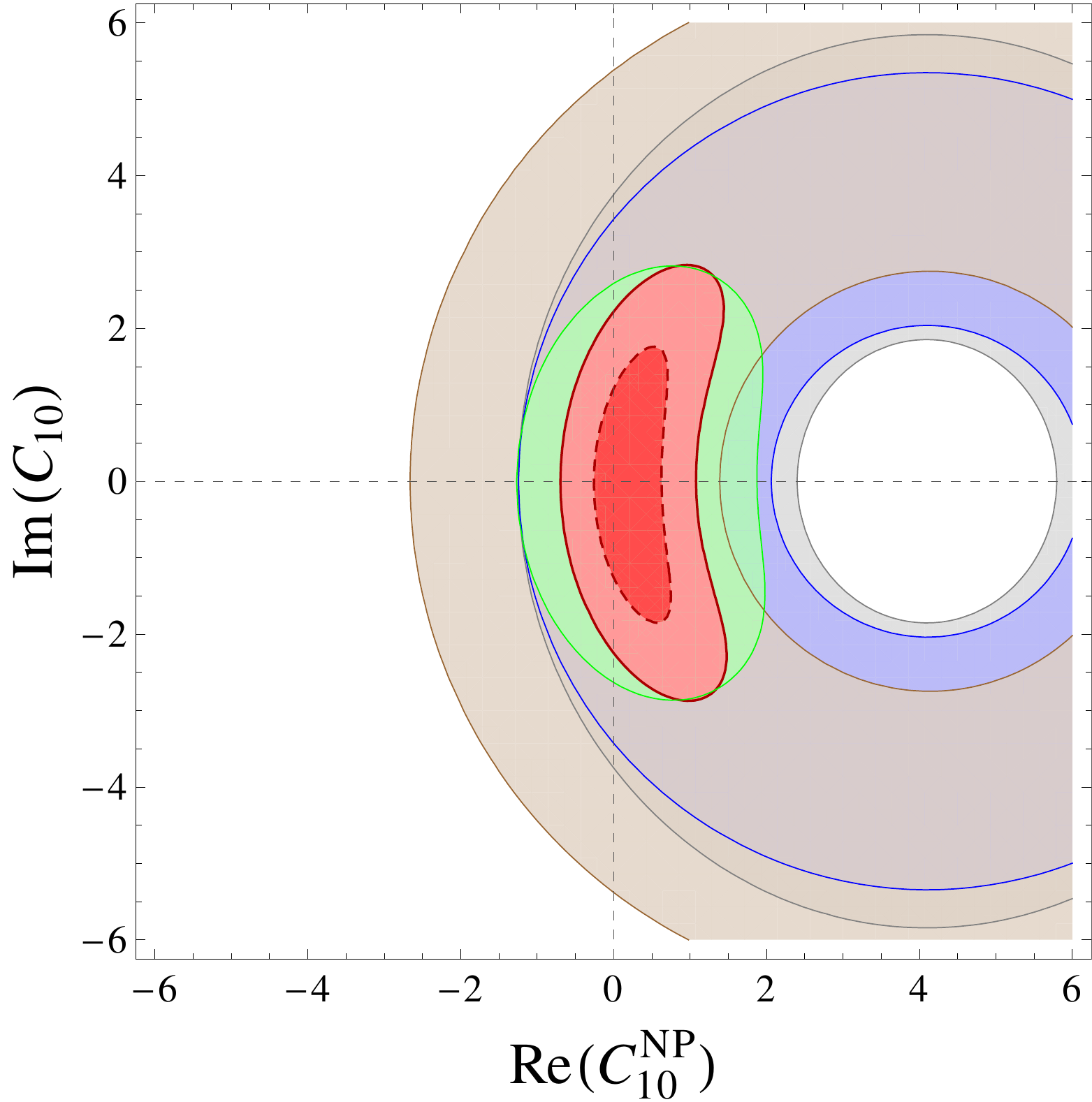} ~~~~~~
\includegraphics[width=.46\textwidth]{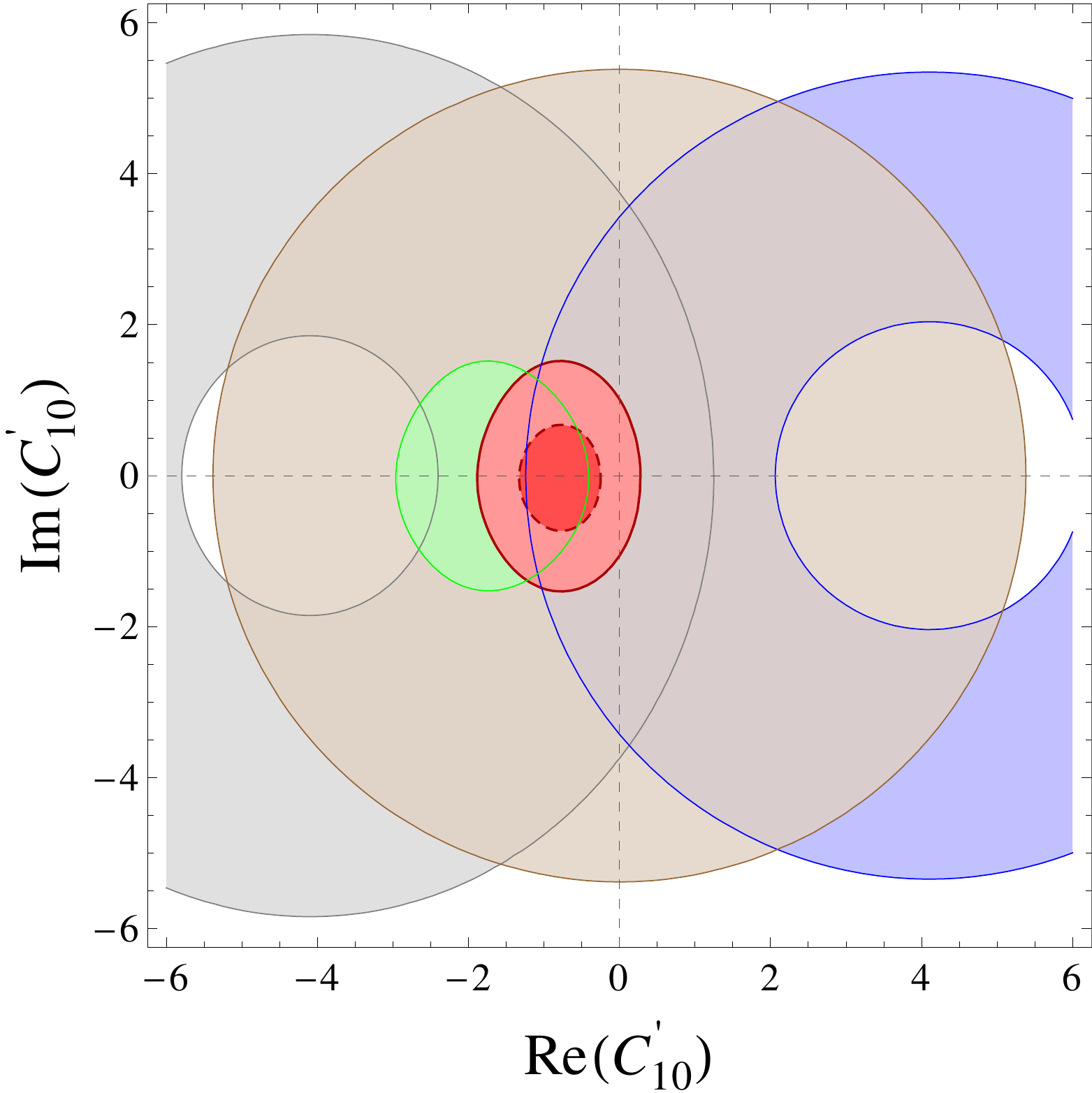}
\caption{Individual $2\sigma$ constraints on the NP contributions to the Wilson coefficients $C_{10}$ (left plot) and $C_{10}^\prime$ (right plot) from $B\to X_s\ell^+\ell^-$ (brown), $B\to K^*\mu^+\mu^-$ (green), $B\to K\mu^+\mu^-$ (blue) and $B_s\to\mu^+\mu^-$ (gray) as well as combined $1\sigma$ and $2\sigma$ constraints (red). (Update of~\cite{Altmannshofer:2012az}, from~\cite{Straub:2013uoa}.)
}
\label{fig:bandplots1}
\end{figure}

The coefficients $C_{10}$ and $C_{10}^\prime$ can be constrained by measurements of $B_s\to\mu^+\mu^-$, $B\to X_s\ell^+\ell^-$, $B\to K^*\mu^+\mu^-$, and $B\to K\mu^+\mu^-$ (see~\cite{Altmannshofer:2011gn,Beaujean:2012uj,DescotesGenon:2012zf,Altmannshofer:2012az} for recent works). In particular, the recent LHCb measurement of the $B\to K^*\mu^+\mu^-$ angular observables $S_3$ and $A_9$~\cite{Aaij:2013iag}, leads to stringent constraints on the right-handed coefficient $C_{10}^\prime$. The constraints on NP contributions to $C_{10}$ and $C_{10}^\prime$ are summarized in Figure~\ref{fig:bandplots1}. 

The obtained bounds on the individual Wilson coefficients can also be translated into bounds on the new physics scale $\Lambda$ that suppresses flavor violating dimension 6 operators in the $b \to s$ effective Hamiltonian $\mathcal{H}_{\rm eff} = \mathcal{H}_{\rm eff}^{\rm SM} + \sum_i c_i \mathcal{O}_i / \Lambda^2$. Assuming generically $|c_i| = 1$, new physics in the semi-leptonic operators $\mathcal{O}_{10}$ and $\mathcal{O}_{10}^\prime$ is probed up to several 10's of TeV. The scalar-operators $\mathcal{O}_S - \mathcal{O}_S^\prime$ and $\mathcal{O}_P - \mathcal{O}_P^\prime$ are probed up to scales of $\Lambda \simeq 100$~TeV~\cite{Altmannshofer:2012az}. 
Considering the current bound on BR$(B_d \to \mu^+\mu^-)$, we find that the corresponding scalar operators in the $b \to d$ sector are already constrained up to scales of $\Lambda \simeq 200$~TeV.

\subsection{$\mathbf{B_s \to \mu^+\mu^-}$ vs $\mathbf{B_d \to \mu^+\mu^-}$}

As is well known, the measurement of both BR$(B_s \to \mu^+\mu^-)$ and BR$(B_d \to \mu^+\mu^-)$ gives a very clean probe of new sources of flavor violation beyond the CKM matrix.
Indeed, in models with Minimal Flavor Violation (MFV), the ratio of the 2 branching ratios is mainly determined by the ratio of the corresponding CKM matrix elements~\cite{Buras:2003td}
\begin{equation}
\frac{{\rm BR}(B_s \to \mu^+ \mu^-)}{{\rm BR}(B_d \to \mu^+ \mu^-)} \propto \frac{f_{B_s}^2}{f_{B_d}^2} \frac{\tau_{B_s}}{\tau_{B_d}} \frac{|V_{ts}|^2}{|V_{td}|^2} ~.
\end{equation}
This relation holds in the SM, models with MFV, but also in models with minimally broken $U(2)^3$ flavor symmetry~\cite{Barbieri:2012uh}. 
Given the measured value of ${\rm BR}(B_s \to \mu^+ \mu^-)$, one can obtain upper and lower bounds on the corresponding $B_d$ branching ratio in these classes of models. We find at $2\sigma$
\begin{equation}
0.3 \times 10^{-10} \lesssim {\rm BR}(B_d \to \mu^+ \mu^-) \lesssim 1.8 \times 10^{-10} ~.
\end{equation}
A measurement of ${\rm BR}(B_d \to \mu^+ \mu^-)$ outside the above range would be a clear indication of new sources of flavor violation beyond the CKM matrix. It would rule out not only the Standard Model but all models with MFV and models with a minimally broken $U(2)^3$ flavor symmetry.

\subsection{Implications of the BR$\mathbf{(B_s \to \mu^+\mu^-)}$ Measurement for Models of New Physics}

The recent experimental results on the BR$(B_s \to\mu^+\mu^-)$ have been interpreted in many models of NP. Among them are the Minimal Supersymmetric Standard Model (MSSM)~\cite{Altmannshofer:2012ks,Arbey:2012ax,Kowalska:2013hha}, models with partial compositness~\cite{Straub:2013zca}, generic 2 Higgs doublet models~\cite{Crivellin:2013wna}, models with additional vector-like fermions~\cite{Buras:2013td}, as well as setups with flavor changing $Z$ or $Z^\prime$ boson~\cite{Buras:2012jb,Guadagnoli:2013mru} and flavor changing scalars~\cite{Buras:2013rqa}.
Here, we briefly discuss the implications of the BR$(B_s \to\mu^+\mu^-)$ measurement on the MSSM. 

It is well know that in the MSSM with large $\tan\beta$, order of magnitude enhancements of the $B_s \to\mu^+\mu^-$ branching ratio are in principle possible, even assuming MFV~\cite{Choudhury:1998ze,Babu:1999hn}.
In the large $\tan\beta$ regime, the most important SUSY contributions come from so-called Higgs penguin diagrams that, assuming MFV, are mainly induced by Higgsino-stop loops and contribute to the Wilson coefficients $C_S$ and $C_P$. They scale with $(\tan\beta)^3$ and do not decouple with the superpartner masses but with the mass of the pseudo-scalar Higgs boson $M_A$. The main parameter dependence is given by
\begin{equation}
 C_S^{\tilde H} \simeq - C_P^{\tilde H} \propto \frac{y_t^2}{16\pi^2} \frac{\mu A_t}{m_{\tilde t}^2} ~\frac{(\tan\beta)^3}{M_A^2} (V_{tb} V_{ts}^*)~.
\end{equation}
The sign of the SUSY contribution is set by the relative sign of the Higgsino mass $\mu$ and the stop trilinear coupling $A_t$.

It is worth mentioning that in the limit $C_S = - C_P$, there exists a model-independent lower bound on the branching ratio~\cite{Altmannshofer:2012ks,Buras:2013uqa}
\begin{equation} \label{eq:bound}
{\rm BR}(B_s \to \mu^+\mu^-) \geq \frac{1}{2}(1-y_s) \times \overline{\rm BR}(B_s \to \mu^+\mu^-)_{\rm SM} ~.
\end{equation}
Therefore the current experimental {\it lower bound} on the $B_s \to \mu^+\mu^-$ branching ratio does not lead to any constraints on the parameter space of the considered MSSM framework, yet. 

\begin{figure}[tb]
\centering
\includegraphics[width=0.46\textwidth]{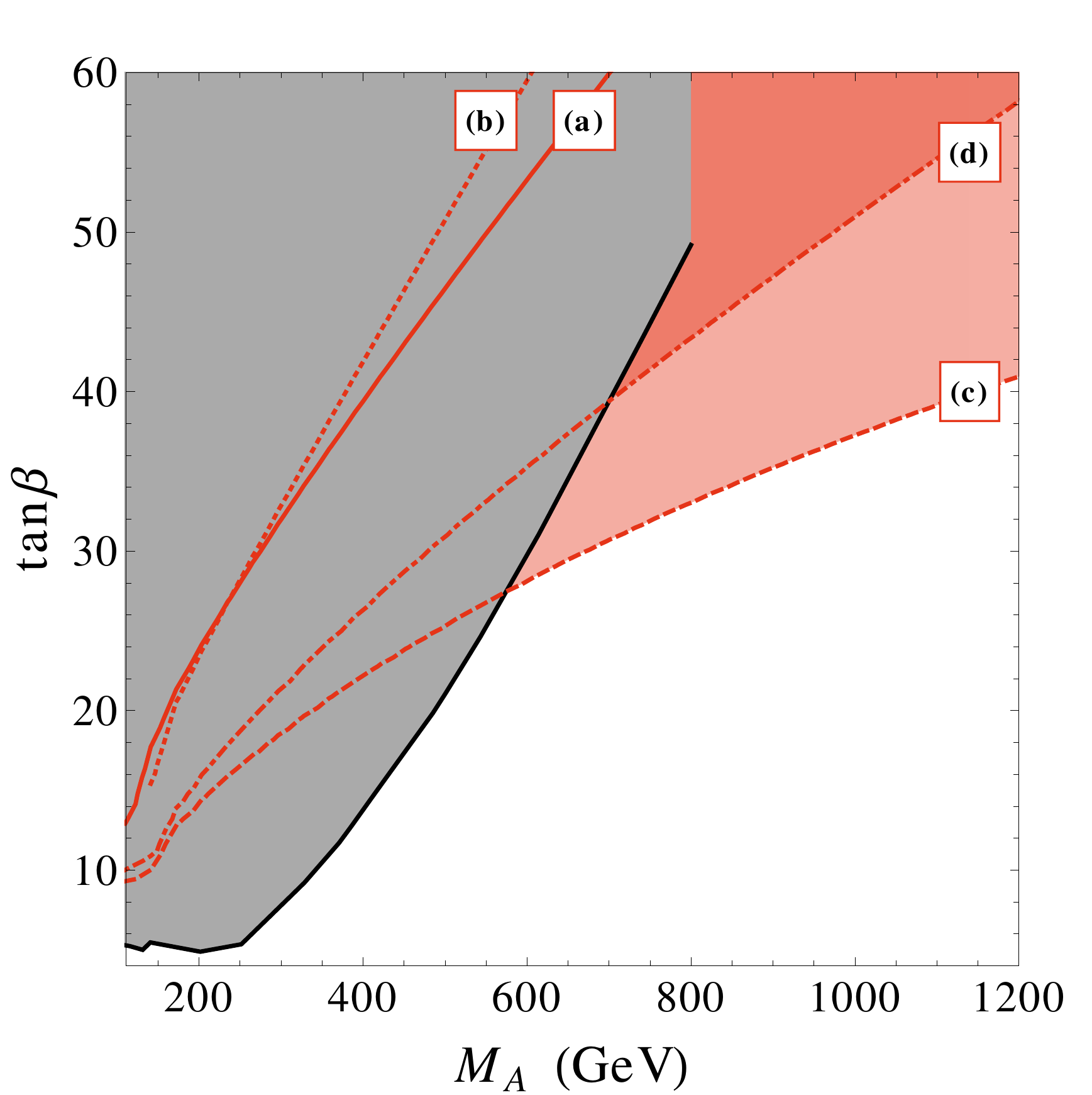} ~~~~~~
\raisebox{100pt}{
\begin{minipage}{0.46\textwidth} \small
\renewcommand{\arraystretch}{1.7}
\renewcommand{\tabcolsep}{6pt}
\begin{center}
\begin{tabular}{|c|cccc|}
\hline
Scenario & (a) & (b) & (c) & (d) \\
\hline\hline
$\mu$ (TeV)& 1 & 4 & -1.5 & 1 \\
sign($A_t$) & + & + & + & -  \\
\hline
\end{tabular}
\end{center}
\end{minipage}}
\caption{Constraints in the $M_A$--$\tan\beta$ plane from the $B_s \to
  \mu^+ \mu^-$ decay in the MSSM with MFV. The red solid, dotted, dashed and dash-dotted
  contours correspond to scenarios (a), (b), (c) and (d), as defined in the table on the right-hand side.  
  All squark soft masses are set to a common value $m_{\tilde q} = 2$~TeV. The magnitude of the stop trilinear coupling $A_t$ is adjusted such that the light Higgs mass is $M_h = 125$~GeV.
  The gray region is excluded by direct searches of MSSM
  Higgs bosons in the $H/A \to \tau^+ \tau^-$ channel. (From~\cite{Altmannshofer:2012ks}.)}
\label{fig:MSSM}
\end{figure}

The constraints from the experimental {\it upper bound} on the branching ratio are illustrated in Figure~\ref{fig:MSSM} in the plane of the pseudo-scalar Higgs mass $M_A$ and $\tan\beta$, for exemplarily chosen values for the soft SUSY breaking parameters and the Higgsino mass.
We observe that the current measurement of BR$(B_s \to \mu^+\mu^-)$ does lead to strong constraints in the MSSM even in the restrictive case of MFV as long as the SUSY contributions interfere constructively with the SM amplitude (as in scenarios (c) and (d)). The $B_s \to \mu^+\mu^-$ constraints extend far above the current limits from direct searches, that stop at $M_A = 800$~GeV. Once the experimental lower bound will be improved above the model-independent bound in~(\ref{eq:bound}), also destructively interfering SUSY contributions (as in scenarios (a) and (b)) will be constrained comparably. Note, however, that the BR$(B_s \to \mu^+\mu^-)$ constraints can be avoided in regions of parameter space with low or moderate $\tan\beta$ and large $M_A$. 

\section{Conclusions}

The recent measurement of the $B_s \to \mu^+\mu^-$ branching ratio by LHCb excludes spectacular new physics effects in this rare decay. On the other hand, rather sizable NP contributions of O(50\%) are starting to be probed only now. As the $B_s \to \mu^+\mu^-$ branching ratio can be predicted with high accuracy in the SM, even moderate NP effects can be clearly identified with increased statistics. Even if the branching ratio will turn out to be completely SM-like, additional observables, like the effective $B_s \to \mu^+\mu^-$ life time could reveal NP in $B_s \to \mu^+\mu^-$.

In the case of $B_d \to \mu^+\mu^-$, NP enhancements by almost an order or magnitude are still allowed by the present data. A measurement of the $B_d \to \mu^+\mu^-$ branching ratio more than a factor of 2 above its SM prediction would not only rule out the SM, but would also imply the existence of new sources of flavor violation beyond the CKM matrix.

In the absence of any direct sign of NP at the LHC, indirect probes of NP, like the rare decays $B_s \to \mu^+\mu^-$ and $B_d \to \mu^+\mu^-$ are more important than ever. In the context of concrete NP models, like the MSSM with MFV, they can exclude regions of parameter space that are currently not covered by direct searches. Model independently, the $B_s \to \mu^+\mu^-$ and $B_d \to \mu^+\mu^-$ decays probe already generic NP at scales of 100 TeV and above, far beyond the direct reach of the LHC.

\section*{Acknowledgments}
It is a pleasure to thank the organizers of the Beauty 2013 conference for the kind invitation to give this talk. I would like to thank David Straub for useful discussions and Stefania Gori for a reading of the manuskript. Fermilab is operated by Fermi Research Alliance, LLC under Contract No.
De-AC02-07CH11359 with the United States Department of Energy.  



\end{document}